\documentclass[]{JHEP3}
\input epsf
\usepackage{epsfig}

\usepackage{bm}

\def\centerbox#1#2{\centerline{\epsfxsize=#1\textwidth\epsfbox{#2}}}
\def\Tr{\,{\rm Tr}\:}
\def\commut#1#2{\Big[#1\,,\, #2\Big]}
\def\be{\begin{equation}}
\def\ee{\end{equation}}
\def\bea{\begin{eqnarray}}
\def\eea{\end{eqnarray}}

\def\Eq#1{Eq.~(\ref{#1})}

\def\O{{\cal O}}
\def\vs{v_{\rm s}}

\def\x{{\bm x}}
\def\p{{\bm p}}
\def\q{{\bm q}}
\def\r{{\bm r}}
\def\k{{\bm k}}
\def\v{{\bm v}}

\def\naBla{{\bm \nabla}}

\def\da{d_{\rm A}}

\def\nc{N_{\rm c}}

\def\mD{m_{\rm D}}

\newcommand\ansatz{{\it Ansatz}}

\def\alphas{\alpha_{\rm s}}

\def\gsim{\mbox{~{\raisebox{0.4ex}{$>$}}\hspace{-1.1em}
        {\raisebox{-0.6ex}{$\sim$}}~}}
\def\lsim{\mbox{~{\raisebox{0.4ex}{$<$}}\hspace{-1.1em}
        {\raisebox{-0.6ex}{$\sim$}}~}}

\advance\parskip 2pt

\title
    {
      Bulk viscosity and spectral functions in QCD
    }

\author{Guy D.~Moore and Omid Saremi \\
    Department of Physics,
    McGill University, 
    3600 University St.,
    Montr\'{e}al, QC H3A 2T8, Canada
    }%
\date {February 2003}

\begin {abstract}%
    {%
      We examine the behavior of the spectral function for the $T^\mu_\mu$
      operator in QCD in the two regimes where it is possible
      to make analytical progress; weak coupling, and close to a second
      order QCD phase transition.  We determine the behavior of the bulk
      viscosity in each regime.  We discuss the problem of analytic
      continuation of the (lattice) Euclidean correlation function to
      determine the spectral function.  In each case the spectral
      function has a narrow peak at small frequency; its shape would be
      challenging to extract accurately from lattice data with error bars.
    }%
\end {abstract}

\begin {document}

\section {Introduction}

Ongoing experiments at RHIC are exploring the QCD plasma, achieving
initial temperatures which exceed the QCD phase transition (or
crossover) temperature.
Since the ions are large compared to the
intrinsic QCD scale, they create a ``macroscopic'' sample of plasma,
which allows the exploration of QCD at these temperatures in its
hydrodynamic regime.  The experimental program indicates that at central
rapidities the plasma indeed behaves hydrodynamically, developing rather
large radial and elliptic flows \cite{RHIC_xpt}.  Viscous hydrodynamic
analyses \cite{nonideal_hydro} only describe the observed elliptic
flow well if the viscosity is very small, $\eta/s < 0.3$.  Radial
flow is caused by outward pressure accelerating the outer layers of the
``fireball'' radially.  It builds up through the whole history of the
fireball expansion until the hadrons become too dilute to interact.  It
is therefore sensitive to the whole time history of the plasma after the
collision.  It is sensitive to the equation of state, since this sets
the relation between pressure and energy density.  It should also be
sensitive to bulk viscosity.  The definition of bulk viscosity is a drop
in the pressure, relative to the equilibrium value at the same energy
density, due to expansion:
\be
P = P_{\rm equil.} - \zeta \,\naBla {\cdot} \v
\ee
with $\v$ the flow velocity and $\zeta$ the bulk viscosity (and with all
variables measured in the instantaneous local rest frame).  Since the
bulk viscosity reduces the outward pressure, it lowers the amount of
radial flow.  Of course, under a single set of experimental conditions
the bulk viscosity may approximately mimic a modification in the 
equation of state; but comparing collisions of different centralities or
of different sized nuclei,
for which $\naBla {\cdot} \v$ will vary, it should in principle be
possible to isolate bulk viscosity from the equation of state.

At high temperatures where the coupling is weak and the theory is nearly
conformal, the bulk viscosity is expected to be small
\cite{HosoyaKajantie,Jeon,ADM}.  Near the phase transition or crossover,
however, it may be appreciable.  This has encouraged a recent reanalysis
of the bulk viscosity in QCD.  In particular, Kharzeev and Tuchin have
recently argued \cite{Kharzeev} that the bulk viscosity can be
determined from lattice results for the equation of state (particularly
the temperature dependence of the pressure $P$ and energy density
$\epsilon$) by using an exact sum
rule and an \ansatz\ for the functional form of the spectral function of
stress-stress correlations.
And Meyer \cite{Meyer} has performed a (pure-glue) lattice study of the
Euclidean $T^\mu_\mu$ correlation function, with a view towards an
analytic continuation of the lattice data to determine the real-time
spectral function and in particular the bulk viscosity.

Using Euclidean data to reconstruct real-time correlation functions is
in general ill-posed without some assumptions about the shape of the
real-time correlation function.  Therefore we feel that it is useful to
learn whatever we can about the spectral function for $T^\mu_\mu$
in whatever regimes analytic information is
available.  We have found two such regimes.  At weak coupling (high
temperature) we can compute the spectral function perturbatively.  And
near a second order phase transition (such as probably occurs in
realistic QCD at some point in the $T$--$\mu$ plane), we can make
reliable statements about scaling behaviors based on universality
arguments.  The latter case may be of importance for real-world QCD; if
the critical point is close enough to the temperature axis, heavy
ion collisions may explore a near-critical crossover, with long
correlation lengths and sensitive $T$ dependence of thermodynamical
variables.

As discussed above, the bulk viscosity is defined as a deviation of the
pressure $P = \frac{1}{3} T_i^i$ from its equilibrium value due to
expansion.  The operator which generates expansion of the system is also
$\frac{1}{3} T_i^i$, and as Kubo showed, one can treat a slowly
expanding system by coupling this operator to an external source.  The
bulk viscosity is determined as the linear response of the operator
$\frac{1}{3} T_i^i$ to such an external source
\cite{Kubo}:
\be
\zeta = \frac{1}{2} \lim_{\omega \rightarrow 0^+} \frac{1}{\omega}
\int_{-\infty}^{\infty} dt e^{-i\omega t} \int d^3 x
\left\langle \commut{{\textstyle \frac{1}{3} T_i^i}(x,t)}
{{\textstyle \frac{1}{3} T_i^i}(0,0)} \right\rangle \,.
\ee
Because the energy $\int d^3 x T^{00}$ is conserved, it is harmless to
shift $T_i^i$ by the energy or any multiple of the energy in the above.
Two useful choices are $k$ such that $\langle T_i^i+kT_0^0 \rangle=0$
(so the operator we use has vanishing expectation value) and $k=-1$, so
the correlator is replaced with a correlator of $T_\mu^\mu$.%
\footnote{We use [${-}{+}{+}{+}$] metric conventions.}

The purpose of this paper is to analyze the spectral function
\be
\rho(\omega) \equiv \int dt e^{-i\omega t} \int d^3 x \frac 19 \left\langle
\commut{T^\mu_\mu(x,t)}{T^\nu_\nu(0,0)} \right\rangle
\label{eq:rho}
\ee
both in the weak coupling regime and close to the phase transition
point, focusing on low frequencies.  The bulk viscosity is
determined by $\zeta = \frac{1}{2} \lim_{\omega \rightarrow 0}
\rho(\omega)/\omega$.  Section \ref{sec:perturbative} examines
the perturbative regime, generalizing the bulk viscosity calculation
of \cite{ADM} to nonzero frequencies.  We show that $\rho/\omega$ has
a peak at zero frequency of height $\O(\alphas^2 T^3)$ and area 
$\int d\omega \rho/\omega \sim \alphas^{7/2} T^4$.  Both the width
and area under the peak contradict the Kharzeev-Tuchin study
\cite{Kharzeev}.
Section \ref{sec:phasetransition} studies the spectral function and bulk
viscosity near a second order phase transition.  Drawing on work by
Onuki \cite{Onuki}, we argue that the bulk viscosity shows a {\em
  power divergence} as the critical point is approached; along the
crossover line a distance $t$ in the $T,\mu$ plane from the
critical point, $\zeta \propto t^{-z\nu+\alpha}$, with $\nu=.630$
the scaling exponent for the correlation length, $\alpha=.110$ the
critical exponent for the heat capacity, and $z\simeq 3$ the dynamical
critical 
exponent.  Again, this corresponds to a sharp peak in $\rho/\omega$ near
zero frequency.  This result contradicts Ref.\ \cite{Kharzeev2}.
We end with a discussion section in which we examine the implications of
our results for the program of determining the spectral function via
analytic continuation of Euclidean data.

\section{spectral function at weak coupling}
\label{sec:perturbative}

For simplicity we will consider only pure glue QCD here; the behavior of
QCD with quarks of negligible mass is more complicated but qualitatively the
same. The trace of the stress tensor is the generator of dilatations,
which are a classical symmetry of the theory.  However this symmetry is
broken at the quantum level: under dilatations the action
$S=\int d^4 x \frac{1}{2g^2} \Tr G_{\mu\nu} G^{\mu\nu}$ changes because
the inverse gauge coupling $1/2g^2$ changes by $-\beta/g^4$,
with $\beta = \frac{\mu^2 d}{d\mu^2} g^2 \sim g^4$ the beta function for the
coupling $g^2$.  Therefore $T^\mu_\mu$ can be replaced with
$T^\mu_\mu \rightarrow \frac{\beta}{g^4} \Tr G^2$ the square of the
field strength \cite{anomaly}.
The correlation function \Eq{eq:rho} is therefore $4\beta^2/g^4$ times
the correlation of the Lagrangian density with itself.

The spectral function is related to the Wightman function by a KMS
relation:
\be
G^>(\omega) \equiv \int dt e^{-i\omega t} \int d^3 x
\frac{1}{9} \left\langle T^\mu_\mu(0,0) T^\nu_\nu(x,t) 
\right\rangle_{\rm conn} \, , \qquad
G^>(\omega) = \frac{1}{1-e^{-\omega/T}} \rho(\omega)\,.
\ee
It is convenient to shift the operator $T_\mu^\mu$ by a multiple of
$T_0^0$ such that its expectation value vanishes, so that the connected
correlation function is the same as the full correlation function.
In our case this means that we need to work not in terms of $T_\mu^\mu$
but in terms of $\O \equiv T_\mu^\mu - T^{0}_0 \langle
T_\nu^\nu\rangle/\langle T^0_0\rangle \simeq T_i^i + 3\vs^2 T_0^0$, with
$\vs$ the speed of sound.  This shift to the operator is $\O(g^4)$ and
will only be important for some terms in what follows.
In fact the only difference between using $\O$ and
using $T_\mu^\mu$ is that the $T_\mu^\mu$ correlation function will have
an extra delta function strictly at zero frequency and of height equal
to the energy susceptibility (heat capacity) times
$(\langle T_\mu^\mu \rangle/\langle T_0^0 \rangle)^2$.
The Wightman correlator for
$\O$ is missing this delta function; its value at $\omega=0$ is $2T$
times the bulk viscosity. 

Perturbatively the leading contribution to this correlation
function is that of Figure \ref{fig1}:

\FIGURE{
\centerbox{0.3}{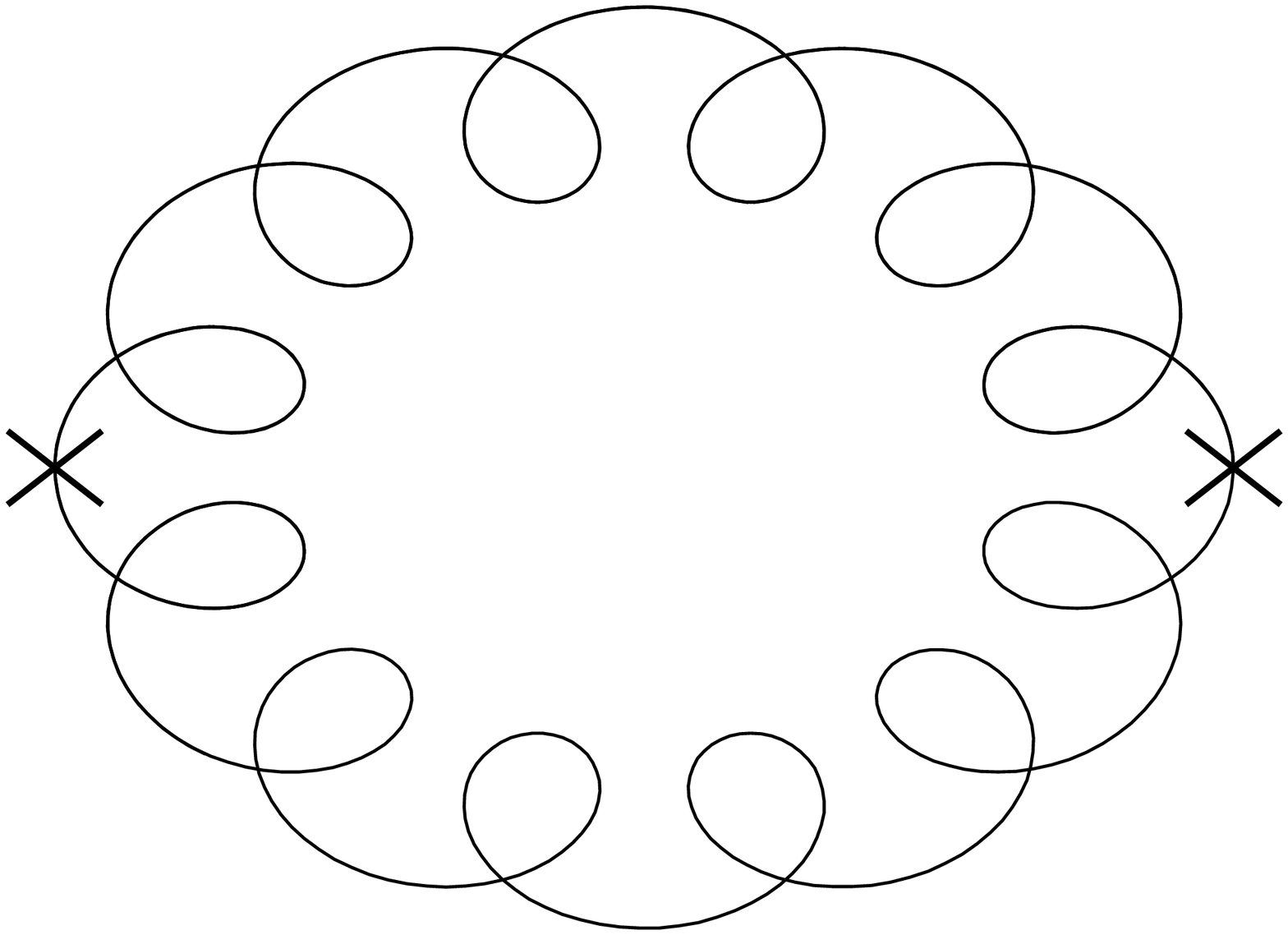}
\caption[one loop graph]
{\label{fig1} Graph which dominates $G^2$--$G^2$ correlations at weak
  coupling.  The $\times$ symbols represent the $G^2$ operator
  insertions.}
}
\noindent
This gives rise to a leading order contribution (defining%
\footnote{Capital letters are 4-vectors; lower case are their spatial
  components.}
$Q=(\omega,0)$) of
\bea
G^>(Q) & = & \frac{2\beta^2(g)\da}{9g^4} \int 
  \frac{d^{4\!} P\, d^{4\!} R}{(2\pi)^4}
  \:\delta^4(Q{-}P{-}R)\;
  G^>_{\mu\alpha\!}(P) G^>_{\nu\beta}(R) 
\nonumber \\ && \hspace{1.3in} \times
   (g^{\mu\nu} P\cdot R {-} P^\mu R^\nu)
    (g^{\alpha\beta} P\cdot R {-} P^\alpha R^\beta) \,,
\eea
where $\da = \nc^2-1$ is the dimension of the group.
Note that both gauge boson propagators are Wightman (cut) propagators,
given by
\be
G^>_{\mu\nu}(P) = [n_b(p^0){+}1]\, 2\pi \delta(P^2{+}m^2)
\sum_{\lambda} \epsilon_\mu(\lambda) \epsilon^*_\nu(\lambda) \,,
\ee
with $\lambda$ the polarization state; $\epsilon_{\mu} P^\mu=0$.  The
use of massive dispersion relations%
\footnote{We use [$-$+++] metric convention}
accounts for forward scattering in the plasma; the plasma mass is
$m^2=\mD^2/2 = g^2\nc T^2/6$.
Since we are working perturbatively, we will treat $\mD^2 \ll T^2$
the energy scale of typical quasiparticle excitations and we will
ignore longitudinal gluons, which have an exponentially small spectral
weight for $p^0 \sim T$ \cite{BraatenPisarski}.

There are two contributions to $G^>(Q)$, corresponding to the two ways the
conditions $\delta^4(Q-P-R) \delta(P^2+m^2) \delta(R^2+m^2)$ can be satisfied
at vanishing $\q$.  Since $|\p|=|\r|$, we have $p^0 = \pm q^0$.  The
contribution where $p^0=r^0=\omega/2$ gives rise to a continuous contribution,
corresponding to a cut in the retarded function:
\be
G^>_{\rm cut} = \left[ n_b({\textstyle \frac{\omega}{2}})+1 \right]^2
   \frac{2\beta^2(g)}{9g^4} \; \frac{2\da\,\omega^4}{32\pi}
  \quad \propto g^4 \omega^4  \,.
\ee
Here $2\da$ is the number of color and spin states of gluons.  For this
contribution, the
difference between $\O$ and $T_\mu^\mu$ is an $\O(g^2)$ correction and
can be neglected.  The parametric behavior is simple to understand:  $g^4$
arises because the trace anomaly makes $T^\mu_\mu$ naturally ${\cal O}(g^2)$
and we are computing the correlator of two $T^\mu_\mu$'s;
and the $\omega^4$ behavior follows on dimensional grounds.  Note that for
$\omega < T$ the behavior changes to $g^4 \omega^2 T^2$ because of the
statistical functions, and is further modified below $\omega \sim gT$.

The other contribution arises when $p^0=-r^0$.  This requires $\omega=0$, and
therefore corresponds to a delta function in the Wightman function (pole in
the retarded function).
If we evaluate the correlation function for $T_\mu^\mu$
operators without subtracting disconnected contributions, we find
\be
G^>_{\rm pole}[\mbox{inc. disconnected}] = \delta(\omega)
\frac{2}{9g^4} \int \frac{p^2 dp}{4\pi E_p^2} \; 2\da\: (\beta\,P^2)^2\:
n_b(p^0) [1{+}n_b(p^0)] \,. 
\ee
This would vanish were it not for dispersion corrections for hard
gluons, mentioned above: $P^2 = -m^2=-\mD^2/2 \sim g^2 T^2$.  Inserting this
estimate and taking $p \sim T$, one finds that the delta-function
contribution naively scales as $g^8 T^5 \delta(\omega)$, with the $g^8$
arising as two powers of $g^2$ from the beta functions and two powers of $g^2$
because of the dispersion relations.

Since this extra $g^4$ suppression may come as a surprise for some readers, we
will pause to discuss its physical origin.
Physically, at zero frequency and momentum the $T^{\mu}_{\mu} \propto
G^{\mu\nu} G_{\mu\nu}$ operator is probing thermal excitations without
disturbing them.  A massless gauge excitation has equal $E^2$ and $B^2$, and
therefore $G^{\mu\nu} G_{\mu\nu}= B^2-E^2$ vanishes for an undisturbed,
propagating gauge boson.  
It is only because of
plasma dispersion corrections, which allow $E^2\neq B^2$, that the
cancellation is not exact.
Therefore the pole contribution ``wants'' to vanish for two reasons; the
smallness of the beta function, contributing $\beta^2/g^4 \sim g^4$, and the
smallness of dispersion corrections, yielding
$(B^2-E^2)^2 \sim (g^2)^2 \sim g^4$.

However there are some complications in evaluating this contribution.  First,
$\beta P^2 = -\beta m^2$ is the same order as $\langle T_\mu^\mu \rangle$.
Therefore the result above is contaminated by disconnected parts, and we should
evaluate the correlation function using the operator ${\cal O}$ defined
above.  This amounts to the substitution
$-\beta m^2 \rightarrow (\vs^2-\frac{1}{3})p^2-\beta m^2$.
The contribution to the Wightman correlator is approximately
\be
\label{eq:pole}
G^>_{\rm pole} = \delta(\omega) \frac{2}{9}
\; 2 \da \frac{1}{4\pi} \int_0^\infty n(p)(1+n(p)) 
\left[ \left(\frac{1}{3}-\vs^2\right) p^2 
        + \frac{\beta \mD^2}{2g^2} \right]^2 \frac{p^2 \:dp}{E_p^2} \,.
\ee
The next complication is that this integral is
infrared singular, indicating that the ``pole''
contribution to the Wightman function is dominated by soft physics.
This happens because $B^2$ and $E^2$ come further from canceling as one
considers more infrared, and therefore more dispersion-corrected,
excitations.  The infrared singular, $\sim \int dp/p^2$, behavior is cut off
at the scale $p \sim gT$, where $E_p^2$ deviates strongly from $p^2$.
A complete treatment of this regime
requires a detailed analysis using Hard Thermal Loop (HTL) effective
theory \cite{BraatenPisarski} and we will not attempt it here.  However
we can easily see that the linear divergence, cut off at the $gT$ scale,
reduces by 1 the power of $g$ appearing in the height of the ``delta
function'' contribution, such that
$\int_{ -gT}^{gT} d\omega G^>(\omega) \sim g^{7} T^5$.
Physically, this is because, in the soft $p\sim gT$ region, there is no
relation between $E^2$ and $B^2$, so the extra $g^4$ suppression found above
is absent for such excitations; however they represent only a $g^3$ fraction
of the energy density, leading to a delta function contribution
$\sim g^3 \times g^4 \sim g^7$ (the $g^4$ still arising from the square of the
beta function).

The last complication is that the ``delta function'' we just found is not
really of zero width; interactions broaden it into a sharp peak.
Since the bulk viscosity is determined by the height of this peak, we
need to determine its shape.  Interactions mean that the delta-function
behavior found above receives corrections.  The delta function arises from
integrating over all (thermal) momenta which can run in the loop in
Fig.~\ref{fig1}.  Very roughly, interactions mean that each particle running
in the loop contributes not a delta function but a Lorentzian of the same
area, with width $\Gamma$ set by the large
angle or large momentum change scattering rate for this particle.
The width $\Gamma$ is momentum dependent and is
parametrically $\Gamma \sim g^4 T^3/p^2$ (see \cite{AMY5} for
a discussion of the relevant scattering processes).  Therefore, although soft
$p\sim gT$ particles dominate the area of the peak, it is hard $p\sim T$
particles which dominate its height, since they have narrower widths.
At frequency $g^2 T \gg \omega \gg g^4 T$ the dominant $p$ is
$p \sim T(g^4 T/\omega)^{1/2}$.  The correlation function is
parametrically
\be
G^>(\omega) \sim \left\{ \begin{array}{ll}
g^4 T^4\,, & \omega \lsim g^4 T \\
g^6 T^4 \sqrt{T/\omega}\,, \qquad & g^2 T \gg \omega \gg g^4 T \\
\end{array} \right. \,.
\ee
For $\omega \gsim g^2 T$ contributions from the Landau cut and higher
order diagrams cannot be neglected.  We will not attempt to address this
region here.

We can make this estimate more precise by extending the results on bulk
viscosity \cite{ADM} (strictly zero $\omega$) to treat $\omega \sim g^4
T$ using the approach of Ref.\ \cite{MooreRobert}.  The idea is that the
correlator $G^>(\omega)$ for $\omega \ll g^2 T$ is essentially
determined by kinetic theory, that is, by solving a Boltzmann equation.
For a detailed diagrammatic justification for this fact (in the context
of scalar field theory) see \cite{Jeon}.  The Boltzmann equation can be
solved variationally using the tools developed in Ref.\
\cite{AMY1} for zero frequency and applied to bulk viscosity in
\cite{ADM}.  Ref.~\cite{MooreRobert} showed how to extend these tools
from zero to small frequency in the context of current-current
correlation functions, and there is no problem doing so for stress
tensor correlation functions too.
The computed shape of the correlator $G^>(\omega) \simeq T
\rho(\omega)/\omega$ for $\omega \sim g^4 T$ is displayed in 
Fig.~\ref{fig2}, which also compares the shape of the ``peak'' in the
$T^\mu_\mu$ correlation function to the peak in the shape of the
$T_{ij}-\frac{\delta_{ij}}{3} T_{kk}$ correlation function relevant for
shear viscosity.  Besides the parametrically large difference in the
heights of the peaks (already discussed above), the figure shows that
the peak in the $T_{ij}$ correlator is much narrower, more closely
resembling a Lorentzian.  This is because, unlike the $T^\mu_\mu$
correlator, the $T_{ij}$ correlator is not sensitive to soft
excitations. 

\FIGURE{
\centerbox{0.7}{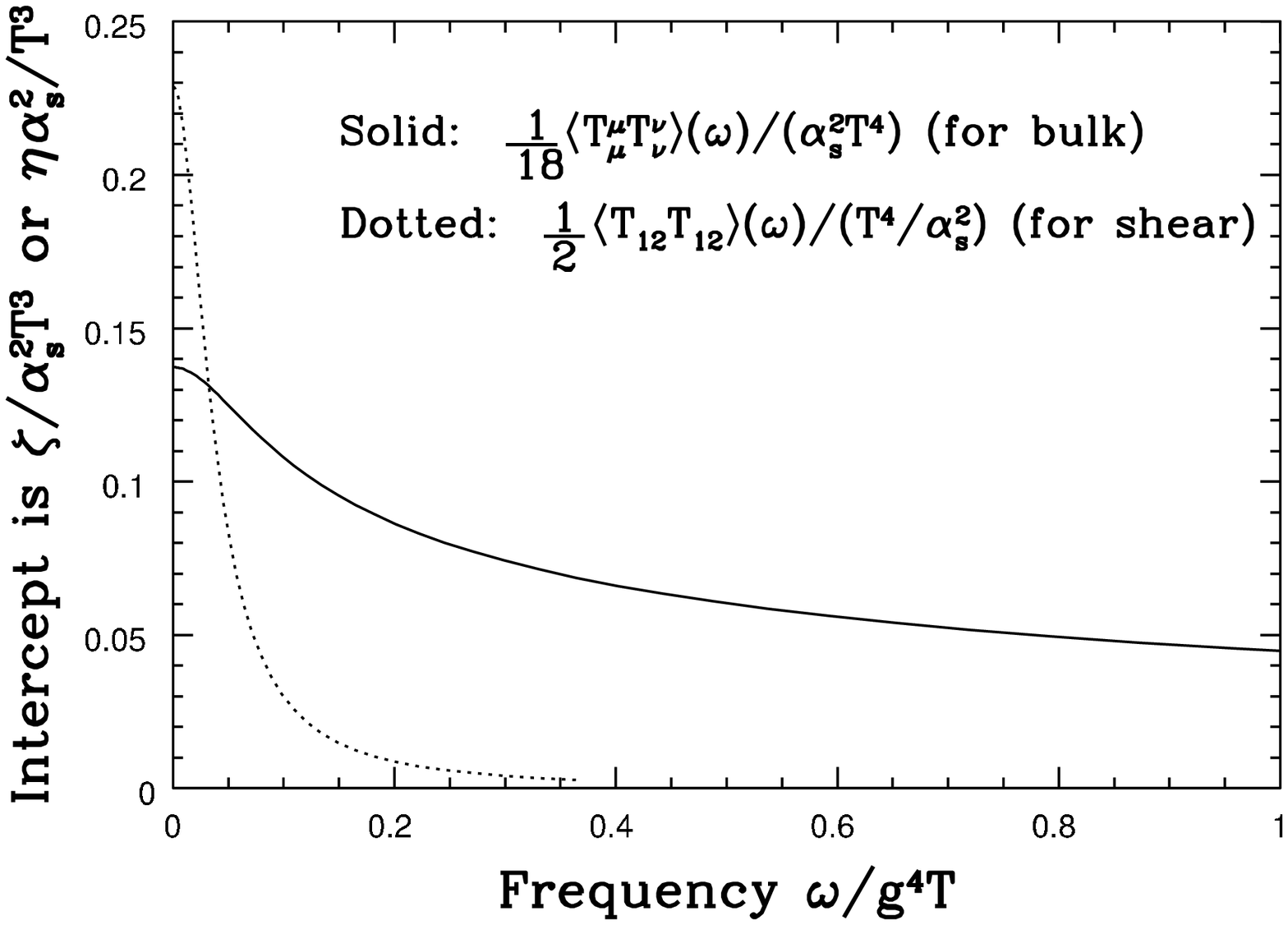}
\caption{\label{fig2}
Wightman correlators at small frequency for the operators
$T^\mu_\mu$ (relevant for bulk viscosity) and 
$T_{ij}-\frac{\delta_{ij}}{3} T_{kk}$ (relevant for shear viscosity) for
$\nc=3$ pure glue QCD at weak coupling (and setting $\mD=1.5T$).  The
``peak'' in the bulk correlation function is wider and has much larger
``shoulders,'' as discussed in the text.
}
}

Let us compare this weak-coupling behavior to that claimed in a recent
analysis by Kharzeev and Tuchin \cite{Kharzeev}.  They derive what they
claim to be an exact sum rule showing that the frequency integral of the
spectral function is related to the temperature dependence of the energy
density and pressure:
\be
\mbox{Eq.\ (12) of \cite{Kharzeev}:} \quad
\int_{-\infty}^\infty \frac{\rho(\omega)}{\omega}d\omega
= T^5 \frac{\partial}{\partial T} \frac{\epsilon - 3 P}{T^4}
+ 16E_{\rm vac} \,.
\label{eq:Kharzeev}
\ee
The vacuum energy contribution can be removed by replacing the spectral
function with its thermal part.  They also state that the perturbative
contributions should be subtracted, though it is not clear to us whether this
refers to the righthand, lefthand, or both sides of the relation.

First we analyze the righthand side.
At weak coupling and in the absence of quark masses it has long been
known that $P$ scales as $T^4$ with a coefficient which can be expanded
in powers of $g$:
\be
P(T) = T^4 \times \Big( A + B g^2[\mu] 
   \left( 1 + \frac{\beta}{g^2} \ln\frac{T^2}{\mu^2} \right) 
   + \O(g^3) \Big)\,.
\ee
The coefficients $A,B$ are known \cite{thermo_NLO} but their precise
values are not important to this discussion.  (In fact the
coefficients through order $g^6 \ln(g)$ are also known
\cite{York}).  All that matters to us here is the functional form and
that the scale dependence of $g^2$ is set by the temperature $T$, which
we have shown explicitly in the above by including the $\O(g^4)$ term
with an explicit log which renders the result $\mu$ independent.
Using that $\epsilon = TdP/dT - P = 3P + 2 T^4 B \beta$, one easily
finds that
\be
\vs^2 = \frac{dP}{d\epsilon} = \frac{1}{3} - \frac{2B}{9A}\beta + \O(g^5)
\sim g^4 \,, \qquad \mbox{and} \qquad
\frac{\epsilon - 3P}{T^4} = 2B \beta  \sim g^4 \,,
\ee
where the coupling $g$ appearing in the expression for $\beta$ should be
understood as being evaluated at $\mu\sim T$.  The dominant $T$
dependence is from this renormalization dependence of $g$:
\be
\frac{Td}{dT} \frac{\epsilon-3P}{T^4} = 2B \frac{Td}{dT} \beta
= \frac{8B}{g^2} \:\beta^2
\sim g^6 \,.
\ee
Therefore a literal interpretation of the righthand side of \Eq{eq:Kharzeev}
leads to a parametrically ${\cal O}(g^6 T^4)$ result.

On the other hand, we can insert our analytical results for the spectral
function into the lefthand side of \Eq{eq:Kharzeev}.  The nearly-delta function
peak gives a contribution of order $g^7 T^4$ (see the discussion after
\Eq{eq:pole}, and use that $\rho/\omega\simeq G^>/T$).  The cut contribution,
after subtracting the vacuum
contribution%
\footnote{
    After vacuum subtraction, the leading-order cut contribution decays
    exponentially.  However, subleading in $g^2$ contributions do not.
    According to S.~Caron-Huot \cite{Caron_private}, the leading
    thermal corrections at large $\omega$ scale as 
    $\rho_T \sim g^2 \rho_{\rm vac} (T/\omega)^4$ both for
    current-current and stress-stress correlation functions, see also
    \cite{Aurenche,Majumder}.
   },
is ${\cal O}(g^4 T^4)$.  
Therefore a literal interpretation of
both sides of the equation leads to an inequality.  And subtracting
the rising cut contribution does not help, since it changes the LHS from
${\cal O}(g^4)$ to ${\cal O}(g^7)$, not $g^6$.  Therefore Kharzeev and
Tuchin's result can only make sense if some subtraction is implied on both
sides of the equation, in which case it has no utility in the perturbative
regime considered here.

\section{spectral function near the 2'nd order transition point}
\label{sec:phasetransition}

There is another regime in which it is possible to say something
analytical about the $T_\mu^\mu$ spectral function.  Close to a second
order phase transition, low frequency and momentum correlation functions
are dominated by long wavelength fluctuations which obey universal
thermodynamical properties which allow for a scaling analysis.
In other words, we can use universality arguments to determine the
functional form of correlation functions, such as their parametric
dependence on the difference between the temperature and the equilibrium
temperature.
Furthermore, hydrodynamic arguments make it possible to extend the
universality predictions to dynamical (unequal time) correlation
functions, though this requires some additional information about what
quantities are conserved \cite{Halperin}.

\FIGURE{
\centerbox{0.4}{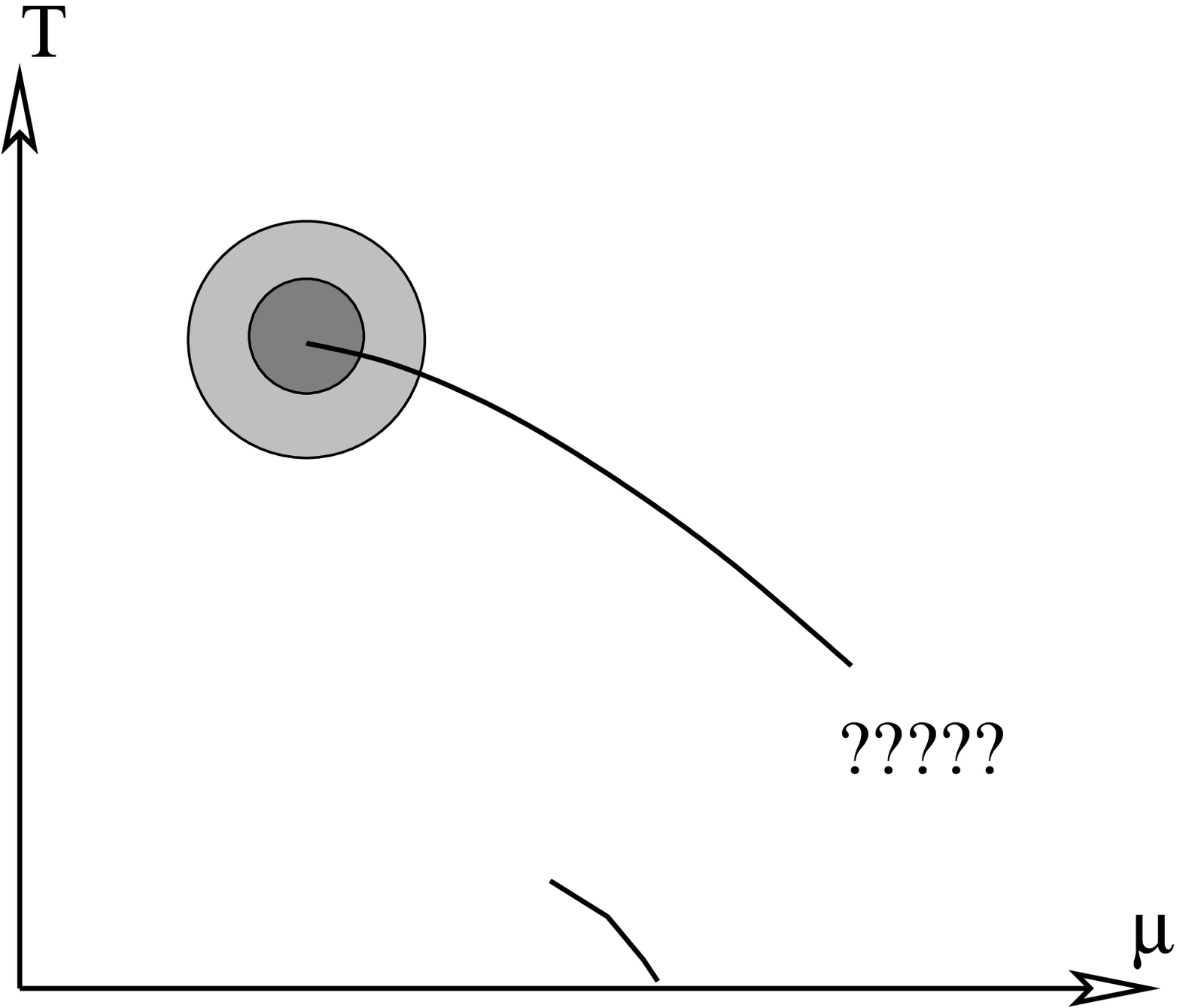}
\caption{\label{fig_latt}  Expected phase structure in the $T-\mu$ plane
  for realistic QCD.  Along the $T$ axis there are no phase transitions,
  but at finite chemical potential there is a 1'st order transition line
  which terminates at an Ising universality class critical point.
  Criticality can give parametric predictions for behavior near this
  point, roughly in the shaded region.  At
  large chemical potential and small temperature there is a 1'st order
  transition associated with nuclei and possibly other transitions
  associated with color superconductivity, which are not important
  here.}
}

It is believed that the phase diagram (in the $T$-$\mu$ plane) for
realistic QCD with two light (but not massless) and
one fairly light quark flavor possesses a first order phase transition
line which terminates in an Ising universality class second order endpoint
\cite{QCD_universal}, as illustrated in Fig.~\ref{fig_latt}
(see however \cite{Philipsen}).
The plasma generated in very high energy heavy ion collisions is
expected to follow a trajectory close to the $T$ axis (small $\mu$),
which probably misses the phase transition line \cite{transition_pt}
but may nevertheless
experience a rather sharp crossover with a
large correlation length.  In principle, intermediate energy heavy ion
collisions may explore the critical point
\cite{how_to_find_critpt}.  In any case, even if it cannot be directly
probed experimentally, it is interesting to consider the vicinity of the
critical point to see what general lessons it teaches us about
$T_\mu^\mu$ correlation functions.  Also note that the pure-glue theory
is expected to have a first-order phase transition point \cite{SY} but
the transition is very weak with a long correlation length
\cite{latt_pureglue} and so scaling arguments might be suggestive here
as well (and two color QCD should have a second order transition in
the Ising universality class \cite{SY}).

Near the critical point, some linear
combinations of the temperature and chemical potential map to the
temperature and magnetic field variables of the Ising model.
Ordinary fluids also have a phase transition between liquid and gas
phases with a critical point in the Ising universality class.  There
is also a mapping of the Ising variables to the temperature and
pressure variables of this system, and therefore a mapping between
$T,\mu$ in QCD and $T,P$ in the liquid-gas system, as illustrated in
Figure \ref{fig_map}.

We need more information to extend universality arguments to unequal
time correlation functions \cite{Halperin}.
The long range fluctuations in the order parameter are coupled to
microscopic degrees of freedom which should lead to diffusive (Langevin)
dynamics for the order parameter fluctuations.  The order parameter
should therefore admit a Ginsburg-Landau
type description, both thermodynamically and dynamically.  This leads as
usual to universality in the behavior of static correlation functions
between all critical systems with the same dimensionality and underlying
symmetries.  But at the level of unequal time dynamics, the
Ginsburg-Landau description must also include any locally conserved
quantities.
In particular, in QCD the temperature and chemical potential are dual to
energy density and baryon number density, which are both conserved.
Since $T$ and $\mu$ map to linear combinations of the Ising variables
$t$ and $h$, QCD behaves like an Ising system with a locally conserved
energy and magnetization.  A local, upward fluctuation in the
magnetization would bias the fluctuations in the order parameter to be
positive in that neighborhood.  Similarly, an upward fluctuation in the
(Ising) energy would bias fluctuations of either sign to be smaller.
These local net modifications would persist as long as the density of
the responsible conserved quantity remained in
that neighborhood.  Therefore there are correlations in the order
parameter which persist as long as the conserved quantities retain
their local values.  But conserved quantities cannot relax locally;
they have to move away, which on large length scales occurs
diffusively.  And the diffusion of these conserved
quantities is in turn sensitive to the order parameter
fluctuations, leading to a coupled
problem.  Fortunately, there is still a notion of universality;
systems with second order phase transitions in the same (static)
universality class, and which have the same set of
conserved quantities, will show the same dynamical scaling behavior
near the transition point \cite{Halperin}.

For realistic QCD the conserved quantities
are the 4-momentum and baryon number.  The dynamical universal behavior
is therefore the same as for the liquid-gas phase transition
\cite{StephanovSon}, since this system also has a conserved energy,
momentum, and particle number.  

\FIGURE{
\centerbox{0.85}{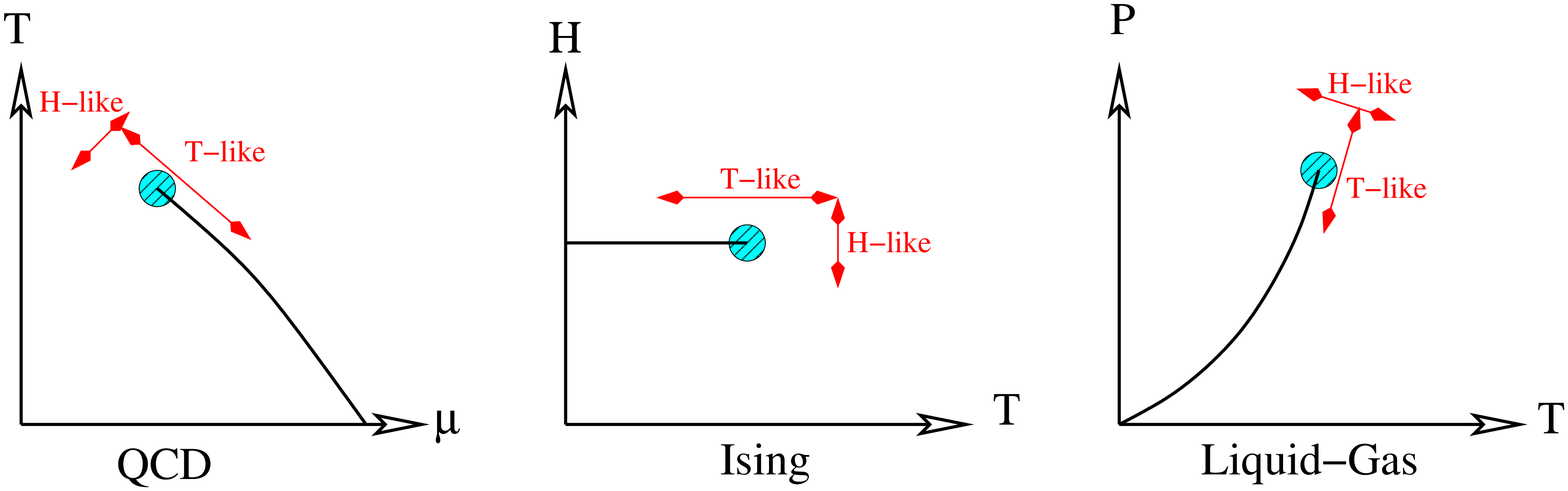}
\caption{\label{fig_map} (Color online) Relation between the Ising model
  temperature and magnetic field directions and the directions in QCD
  and the liquid-gas system.}
}

Fortunately Onuki has performed a detailed study of the dynamical
critical behavior of the liquid-gas system near its critical point,
focusing on the behavior of bulk viscosity \cite{Onuki}.  Since the
dynamical universalities are the same, this can be directly adapted to
the QCD case%
\footnote{%
    It may seem confusing that a relativistic theory, QCD, and a
    nonrelativistic theory, conventional fluids, should display the
    same dynamical criticality.  But remember that at the
    hydrodynamical level, all but one degree of freedom in either
    system evolves dissipatively, so propagation of hydrodynamical
    quantities is slow (nonrelativistic) in either case.  The
    exception is sound waves, which propagate at a fixed velocity
    $c_{\rm s}<1$ in each theory and which turn out not to be
    important to critical behavior \cite{Halperin}.}.
As one
approaches the transition point, the correlation length $\xi$ exceeds
the natural microscopic length scale $l_{\rm mic}$ ($l_{\rm mic}=T^{-1}$
in a relativistic setting).  The dynamics of the order parameter
fluctuations on a scale $l^{-1}_{\rm mic} < k < \xi^{-1}$ are
dissipative and slowly evolving, with
the fluctuations changing on a time scale $\tau_k \propto k^{-z}$,
with $z$ the dynamical critical exponent:
\be
\int d^3 x e^{i\k\cdot \x} \langle \psi(x,\tau) \psi(0,0) \rangle
= \chi(k) e^{-\tau/\tau_k} \, , \qquad \tau_k \sim k^{-z}
\, ,
\ee
with $\psi$ the order parameter and $\chi(k)$ its momentum-dependent
susceptibility.
In 3D Ising systems with liquid-gas dynamical universality, $z \simeq
3$ \cite{Halperin}. Equilibration is dominated by these slow modes.

In the Ising model, the free energy $F$ is a function of the reduced
temperature $t=(T-T_{\rm c})/T_{\rm c}$ and the magnetic field $h$.
Close to the transition point it behaves as
\be
F(t,h) = F_{\rm nonsing}(t,h) 
  + t^{2-\alpha} F_{\rm sing}(t/|h^{1/(\beta{+}\gamma)}|) \,.
\ee
Here $F_{\rm sing}$ is a contribution to the free energy arising from
the long range (near)critical fluctuations.  Note that at small $t$
these fluctuations give almost no contribution to the pressure $P=-F$.
The entropy is $S=-\partial F/\partial t$ and $E=ST+F$.  Therefore the
heat capacity $C_v = \partial E/\partial T$ behaves as
$C_v = C_{v,\rm nonsing} + C_{v,\rm sing} t^{-\alpha}$, displaying a
weak divergence as $t\rightarrow 0$ which arises from the long range
fluctuations in the order parameter.  These fluctuations provide no
pressure but dominate the heat capacity, leading to a speed of sound
$\partial P / \partial \epsilon \simeq 0$.  The same is true in QCD near
the critical point except that the Ising ``temperature'' direction
corresponds to a linear combination of temperature and chemical
potential (so the ``heat capacity'' referred to here is really a
linear combination of heat and baryon number capacity).

A small, rapidly applied compression will not promptly change the
long-range correlations of the order parameter; instead it changes the
noncritical degrees of freedom, and therefore leads to an
instantaneous pressure rise $\Delta P \sim \Delta \epsilon$ typical of
relativistic degrees of freedom.  Then, on a time scale $\tau \sim
\xi^z$, the long range fluctuations equilibrate, absorbing almost all
of $\Delta \epsilon$ (since they dominate the heat capacity) and
allowing the pressure to relax to the equilibrium value.  Therefore
one might guess that $\zeta \sim \xi^z$.  [In the next few paragraphs,
dimensions in parametric estimates are to be filled in with the
appropriate power of the intrinsic scale $T$.]  Along the ``crossover
line'' is the map of the $h=0$, $t>0$ line in the Ising system, this
behavior is $\zeta \sim t^{-z\nu}$, with $t$ the distance in the
$T,\mu$ plane from the critical point and $\nu$ the critical exponent
$\nu\simeq 0.630$ in the Ising system.

In fact a more detailed analysis \cite{Onuki} shows that
$\zeta \sim \xi^{z-\alpha/\nu} \sim t^{-z\nu+\alpha}$.  Onuki has
given a careful derivation of this result; here we give a simple
intuitive explanation of why it is true.  Consider a small, rapid
compression.  After an intermediate amount of
time $\tau$ satisfying $T^{-1} < \tau < \xi^z$, all
modes with $k > \tau^{1/z}$ have equilibrated; those with smaller $k$
(longer wavelength) remain out of equilibrium.  The heat capacity
represented by the equilibrated long-range fluctuations is
$C_{v,k<\tau^{1/z}} \sim k^{\alpha/\nu} \sim \tau^{\alpha / z \nu}$.
Therefore a fraction $\tau^{-\alpha/z\nu}$ of the energy remains in
the noncritical degrees of freedom so the pressure is elevated with
respect to equilibrium by 
$\Delta P \sim \Delta \epsilon \tau^{-\alpha / z \nu}$.  This is true
at all times $\tau < \xi^z \sim t^{-z\nu}$.  The bulk viscosity is
found by integrating this result over all time; it is dominated by
$\tau \sim \xi^z$, and gives $\zeta \sim \xi^{z-\alpha/\nu}$.  Fourier
transforming the $\tau$ dependence gives the spectral function (at
low frequencies); the Wightman function for frequency $\omega$ is
dominated by times $\tau \sim \omega^{-1}$ and is parametrically
$G^>(\omega) \sim \omega^{-1+\alpha/z\nu}$.  Reinserting powers of
$T$, we find
\begin{equation}
G^>(\omega) \sim T^3 (\omega/T)^{-1+\alpha/z\nu} \quad
[\omega > \xi^{-z} T^{1-z}] \, ; \qquad \quad
\zeta \sim T^3 (\xi T)^{z-\alpha/\nu} \,.
\end{equation}
Once again, we find that the spectral function [divided by frequency]
possesses a narrow peak for $\omega < T$.  Note that the total area
under the peak is finite and is essentially determined by noncritical
physics; however the height of the peak diverges as a power of the
distance to the critical point:  $\zeta \sim t^{-z/\nu+\alpha}$ as one
approaches the critical point along the crossover line.

\section{Discussion}
\label{sec:discussion}

In both of the cases where we can gain analytical insight (weak
coupling and close to the second order phase transition point), the
spectral function $\rho(\omega)/\omega$ has a peak at low
frequency.  This peak arises because there are degrees of freedom
which equilibrate very slowly; the width of the peak corresponds to
the inverse of the relaxation time towards equilibrium.  At weak
coupling, all degrees of freedom exhibit slow relaxation.  In this
case relaxation is slowest for high momentum quasiparticles, which
dominate the height of the peak, though low momentum fluctuations
dominate the area under the peak.  Near the phase transition point, it
is long range fluctuations in the order parameter which equilibrate
slowly and give rise to the peak in the spectral function.  These
fluctuations are important to bulk viscosity because
they dominate the heat capacity, though they are of little importance
for shear viscosity since they contribute almost nothing to the
pressure.%
\footnote{%
    Note however that it is believed that shear viscosity also shows a
    weak divergence at the second order transition point for systems
    in the liquid-gas dynamical universality class;
    roughly $\eta \sim t^{-0.05}$ \cite{Halperin}.
    }
The bulk viscosity is determined by the dynamical critical exponent
which determines the critical slowing down of these fluctuations.

It would be too bold to extrapolate from these two examples to claim
that $G^>$ or $\rho/\omega$ always has such a low frequency peak.  For
instance,
at $2T_{\rm c}$, where there are no near-critical fluctuations but the
coupling is strong, it is quite possible that all degrees of
freedom equilibrate quickly and the spectral function is smooth near
zero frequency.  However, the most interesting temperature range
experimentally is temperatures close to $T_{\rm c}$, both because the
plasma almost surely explores such temperatures in real heavy ion
collisions and because it is the only place where we expect the bulk
viscosity to be appreciable.

The presence of a peak in $\rho(\omega)/\omega$ at small $\omega$ is
problematic for the reconstruction of the bulk viscosity from unequal
Euclidean time correlations measured on the lattice.  The Euclidean time
correlation function%
\footnote{
    It is important to compute the connected correlation function on the
    lattice--or equivalently to subtract off the mean value of each
    operator, which is equivalent at nonzero $\tau$ to rescaling the
    $T^{00}$ contribution until the expectation value vanishes.  We
    thank Derek Teaney for emphasizing this point to us.}
\be
G_{\rm E}(\tau) \equiv \int d^3 x \frac{1}{9}
\langle T^{\mu\mu}_{\rm E}(x,\tau) T^{\nu\nu}_{\rm E}(0,0) 
\rangle_{\rm conn}
\ee
is related to the spectral function via the integral relation
\be
G_{\rm E}(\tau) = \int_{-\infty}^{\infty} \frac{d\omega}{2\pi}
  \frac{\rho(\omega)}{\omega} K(\omega,\tau) \, , \qquad
K(\omega,\tau) \equiv \frac{\omega \cosh[\omega(\tau{-}1/2T)]}
                           {\sinh(\omega/2T)} \,.
\ee
In principle, complete knowledge of $G_{\rm E}(\tau)$ as an analytic
function allows for the reconstruction of $\rho(\omega)/\omega$.
However in the lattice context one only has numerical data with error
bars at a finite number of times $\tau$, and some procedure (such as the
Maximal Entropy Method \cite{MEM}) must be used to reconstruct the
spectral function.

\FIGURE{
\centerbox{0.6}{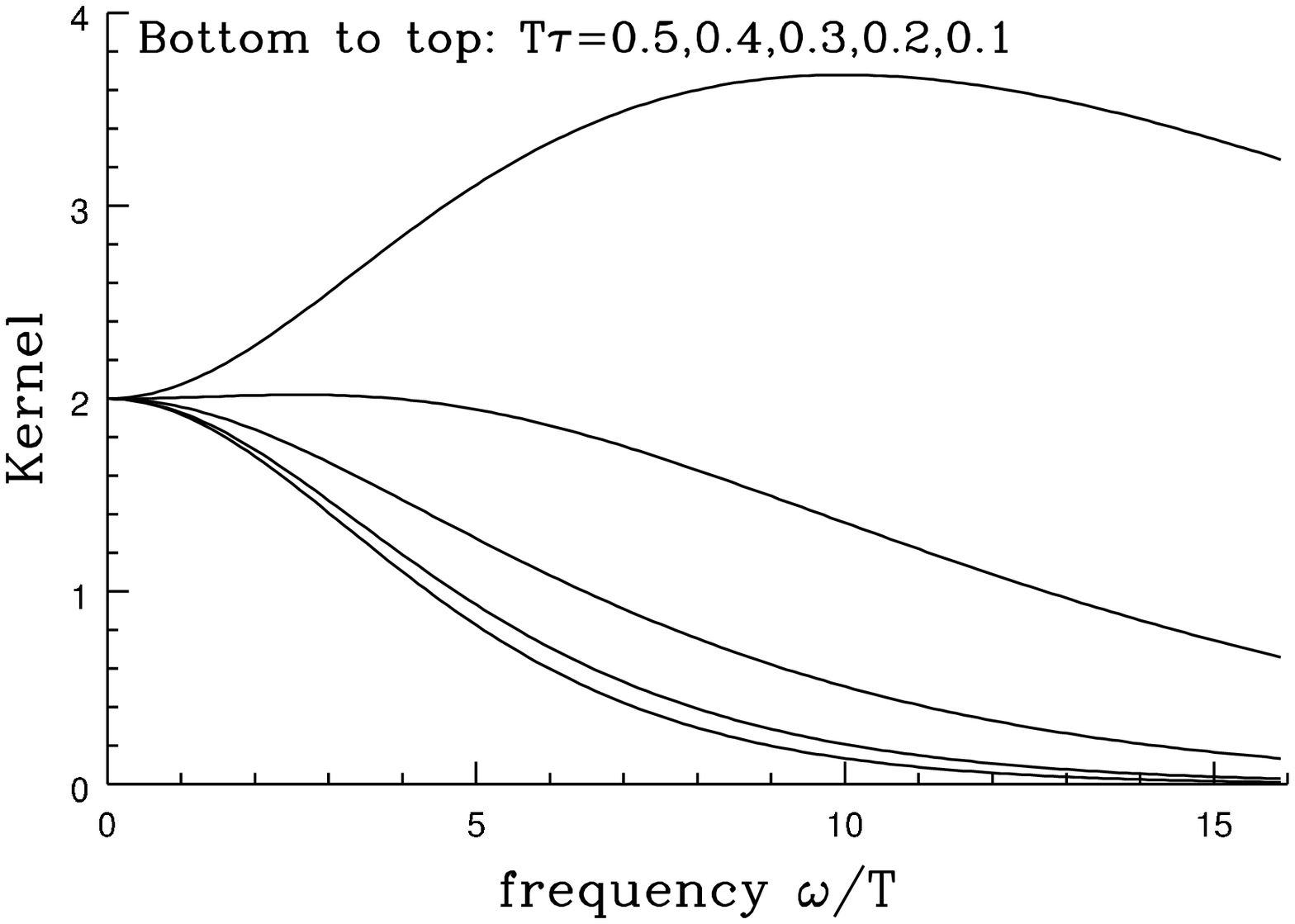}
\caption{\label{fig:K} Kernel $K$ relating the (Minkowski) spectral
  function to the (Euclidean) unequal $\tau$ correlation function, as a
  function of (Minkowski) frequency for some representative values of
  Euclidean time $\tau$.  The behavior at small $\omega$ is almost
  $\tau$ independent.}
}

Figure \ref{fig:K} shows the function $K(\omega,\tau)$ as a function of
$\omega$ for some representative values of $\tau$.  This figure
illustrates the challenge of learning about a sharp spike in the
spectral function:  all of the curves take the same value near zero
frequency.  In other words, each $\tau$ gives degenerate information
about a peak at small frequency--the area under the peak--with almost
no sensitivity to the shape of the peak.  Therefore, details about the
shape of such a peak effectively have to be inserted as part of the
fitting procedure.  Meyer's recent results \cite{Meyer} effectively
assume that $\rho(\omega)/\omega$ is smooth at small $\omega$, which we
have just shown is not a good assumption, at least near a second order
transition or a very weak first order transition.

So what is the best way to estimate $\zeta$ in QCD near the
transition (or crossover) point?  There are two things which we 
{\em can} (in principle) determine reliably from a lattice calculation.  One
is the correlation length $\xi$ for the order parameter
($\langle \bar\psi \psi \rangle$ for nearly chiral 2-flavor QCD).  The
other is the total area under the small momentum peak in the spectral
function.  This can be determined by measuring
$G_{\rm E}(\tau)$ on the lattice and fitting for the spectral function
allowing a narrow peak at the origin, with the area under the peak
used as a parameter in the fit.  Call this area 
$G_{\!\int} = \int_{|\omega|<T} \rho(\omega)d\omega/2\pi \omega$.  [We expect
  it to be almost as large as $G_{\rm E}(\tau/2)$.]
Then we know parametrically that
\be
\zeta = A \frac{G_{\!\int}}{2T} (\xi T)^{z-\alpha/\nu} \,.
\ee
What we do not know is the coefficient $A$.  This coefficient requires
dynamical information; it represents the constant in the scaling
relation between a wave number $k$ and the relaxation time of
fluctuations at that wave number, $T\tau_k = A (T/k)^z$.  To get a
phenomenological estimate of the bulk viscosity where $(\xi T)$ is
large, we have to make some reasonable guess for the coefficient $A$.
We advocate $A=1$ and $A=\pi^z$ as two reasonable choices, based on
the assumptions that the critical regime begins at the scale $T$ and
the scale $\pi T$ respectively.  Clearly there is a rather large band
of uncertainty in the final determined $\zeta$.

In summary, we have shown that in both of the cases where analytic
methods can be brought to bear (weak coupling and close to the second
order transition point), the spectral function needed to
determine the bulk viscosity has a narrow peak at low frequency.  At
weak coupling the peak has a height $\lim_{\omega \rightarrow 0}
\rho(\omega)/\omega \sim g^4 T^4$ and area
$\int d\omega \rho(\omega)/\omega \sim g^7 T^5$; near the critical
point the area is $\sim T^5$ and the 
height diverges as $T^4 (\xi T)^{z-\alpha/\nu}$.  This behavior
complicates the reconstruction of the spectral function from Euclidean
data.

\medskip

\centerline{\bf Acknowledgments}

\medskip

\noindent
We would like to thank Frithjof Karsch, Harvey Meyer, and Derek Teaney
for useful discussions.  This work was supported in part by
the Natural Sciences and Engineering Research Council of Canada and by a
Tomlinson postdoctoral fellowship at McGill University.

\end{document}